\documentclass[journal,twocolumn]{IEEEtran}
\usepackage{amsfonts}
\usepackage{amsmath,amssymb}
\usepackage{acronym}  
\usepackage{algorithm}
\usepackage{algorithmic}
\usepackage{bm}
\usepackage{bbm}
\usepackage{booktabs}
\usepackage{color, soul}
\usepackage{cite}
\usepackage{graphicx}
\usepackage{indentfirst}
\usepackage{lipsum}
\usepackage{setspace}
\usepackage{tikz}
\usetikzlibrary{arrows}
\usepackage{subfigure}
\usepackage[amsmath,thmmarks]{ntheorem}
\usepackage{theorem}
\usepackage{hyperref}
\hypersetup{hidelinks, 
colorlinks=true,
allcolors=black,
pdfstartview=Fit,
breaklinks=true}

\def \H {^{\mathsf{H}}}

\begin{document}

\title{Electromagnetic Information Theory: Fundamentals, Modeling, Applications, and Open Problems}

\author{{Jieao~Zhu, {\textit{Student Member, IEEE}}, Zhongzhichao~Wan, {\textit{Student Member, IEEE}}, Linglong~Dai, {\textit{Fellow, IEEE}},\\
M\'{e}rouane Debbah, {\textit{Fellow, IEEE}}, and H. Vincent Poor, {\textit{Life Fellow, IEEE}} }
\thanks{J. Zhu, Z. Wan, and L. Dai are with the Department of Electronic Engineering, Tsinghua University, Beijing 100084, China, and also with the Beijing National Research Center for Information Science and Technology (BNRist), Beijing 100084, China (e-mails: \{zja21, wzzc20\}@mails.tsinghua.edu.cn, daill@tsinghua.edu.cn).

M. Debbah is with Khalifa University of Science and Technology, P O Box 127788, Abu Dhabi, UAE (e-mail: merouane.debbah@ku.ac.ae). 

H. Vincent Poor is with the Department of Electrical and Computer Engineering, Princeton University, USA (e-mail: poor@princeton.edu).
}}

\markboth{Accepted by IEEE Wireless Communications}%
{Zhu \MakeLowercase{\textit{et al.}}: Electromagnetic information theory: Fundamentals, modeling, applications, and open problems}

\maketitle

\begin{abstract}
	Traditional massive multiple-input multiple-output (MIMO) information theory adopt non-physically consistent assumptions, including white-noised, scalar-quantity, far-field, discretized, and monochromatic EM fields, which mismatch the nature of the underlying electromagnetic (EM) fields supporting the physical layer of wireless communication systems. 
	To incorporate EM laws into designing procedures of the physical layer, we first propose the novel concept of EM physical layer, whose backbone theory is called EM information theory (EIT). 
 	In this article, we systematically investigate the basic ideas and main results of EIT. First, we review the fundamental analytical tools of classical information theory and EM theory. Then, we introduce the modeling and analysis methodologies of EIT, including continuous field modeling, degrees of freedom, and mutual information analyses. Several EIT-inspired applications are discussed to illustrate how EIT guides the design of practical wireless systems. Finally, we point out the open problems of EIT, where further research efforts are required for EIT to construct a unified interdisciplinary theory. 
\end{abstract}


\vspace{-0.5em}
\section{Introduction} \label{sec-intro}
The past decade has witnessed the proliferation of the massive multiple-input multiple-output (MIMO)~\cite{marzetta2010noncooperative} from a theoretical concept to a practical technology. Thanks to the increase of transceiver antennas, the massive MIMO technology has triggered a significant performance improvement in 5G wireless communications. 
However, prevailing analysis and design procedures for massive MIMO are usually based on white-noised, scalar-quantity, far-field~\cite{liu2018novel}, discretized~\cite{goldsmith2003capacity}, monochromatic, and other non-physically consistent assumptions. 
These assumptions will gradually become invalid when an ultradense MIMO, i.e., a continuous-aperture MIMO (CAP-MIMO), is considered. For example, the noise observed at each antenna will exhibit two distinct properties as the number of antennas grows. First, the noise will become correlated due to the strengthened electromagnetic (EM) mutual coupling. Second, the noise power will increase because of the exacerbated thermal fluctuation within small volumes. Both of these two effects will cause the traditional MIMO information-theoretic models to over-estimate the channel capacity, leading to mismatches between the system design and the actual wireless channel properties in the novel architectures beyond massive MIMO. 
In response to these theoretical defects, the designing procedure of traditional wireless physical layer should be extended by re-considering the EM propagation laws, which leads to the novel concept of EM physical layer. 
Correspondingly, a theory that can model and analyze the real-world EM wireless information system with physically interpretable and mathematically reasonable assumptions is of interest, which motivates the research of EM information theory (EIT).

EIT is an interdisciplinary subject that integrates deterministic physical theory and statistical mathematical theory to study information transmission mechanisms in spatially continuous EM fields. 
Specifically, EIT unifies the basic laws and methodologies in both classical EM theory and information theory. 
As a result, EIT is capable of building a framework for system modeling and performance analysis that incorporates EM propagation in the analysis of wireless information systems. 

In this article, we systematically investigate the basics and results of EIT\footnote{Simulation codes are provided to reproduce the results in this paper: \url{http://oa.ee.tsinghua.edu.cn/dailinglong/publications/publications.html}.}. The basic ideas and main results of this article can be summarized as follows:
\begin{itemize}
\item{The fundamental mathematical tools for classical information theory and EM theory are briefly reviewed at first. Specifically, degrees of freedom (DoF), prolate spheroidal wave functions (PSWFs), channel capacity, and Maxwell's equations are introduced. To integrate the probabilistic nature of information theory and the continuous property of EM fields, random fields are then introduced into EIT, which enables performance analysis based on EIT. }
\item{The basic EIT modeling methodologies are introduced for EM channels and EM noise. For EM channel modeling, both the deterministic and stochastic modeling approaches are discussed. For EM noise modeling, the noise fields are first categorized according to their different physical origins, and then the spatial correlation characteristics of different kinds of noise fields are discussed separately. }
\item{EIT performance analysis methods are discussed. By applying the EIT modeling methodologies, the DoF, and the mutual information can be derived using mathematical tools including the PSWFs, Karhunen-Lo\`{e}ve expansion, and random field theories. } 
\item Moreover, several EIT-inspired applications are provided. For example, to approach the EIT capacity, the CAP-MIMO has been proposed by fully harvesting the mutual information in a limited aperture area. Benefiting from the extra spatial DoF predicted by EIT, the location division multiple access (LDMA) technology has been proposed to provide a new possibility for capacity improvement. 
\item{Finally, several open problems of EIT are discussed, including the analog and digital challenges brought by EIT, compatible EM noise modeling, and the evaluation of the EM capacity. We explain the meanings of these open problems, and why they are important for future research in EIT. }
\end{itemize}

\vspace{-1em}
\section{Fundamentals of EIT}
In this section, we first distinguish two different definitions of DoF in information theory, i.e., channel DoF and functional DoF.  
Then, we discuss the concept of channel capacity and its relation to DoF. 
After that, we introduce Maxwell's equations to describe the propagation of EM fields. 
Finally, we employ random fields as the EIT foundation for unifying the probabilistic information theory and deterministic field theory.

\vspace{-1em}
\subsection{Degrees of Freedom}
\label{Sec_2_Subsec_1}
DoF is a mathematical quantity that originally describes the number of independent parameters in a physical system. 
In the communication community, the DoF usually refers to the number of orthogonal subchannels that can independently carry information, which is uniquely determined by the channel structure. It is usually calculated by counting the significant singular values of a given channel matrix~\cite{goldsmith2003capacity}. 
Thus, we use the term {{\emph{channel DoF}}} to describe such a number associated with a channel.

\begin{figure}
	\centering 
	\includegraphics[width=\linewidth]{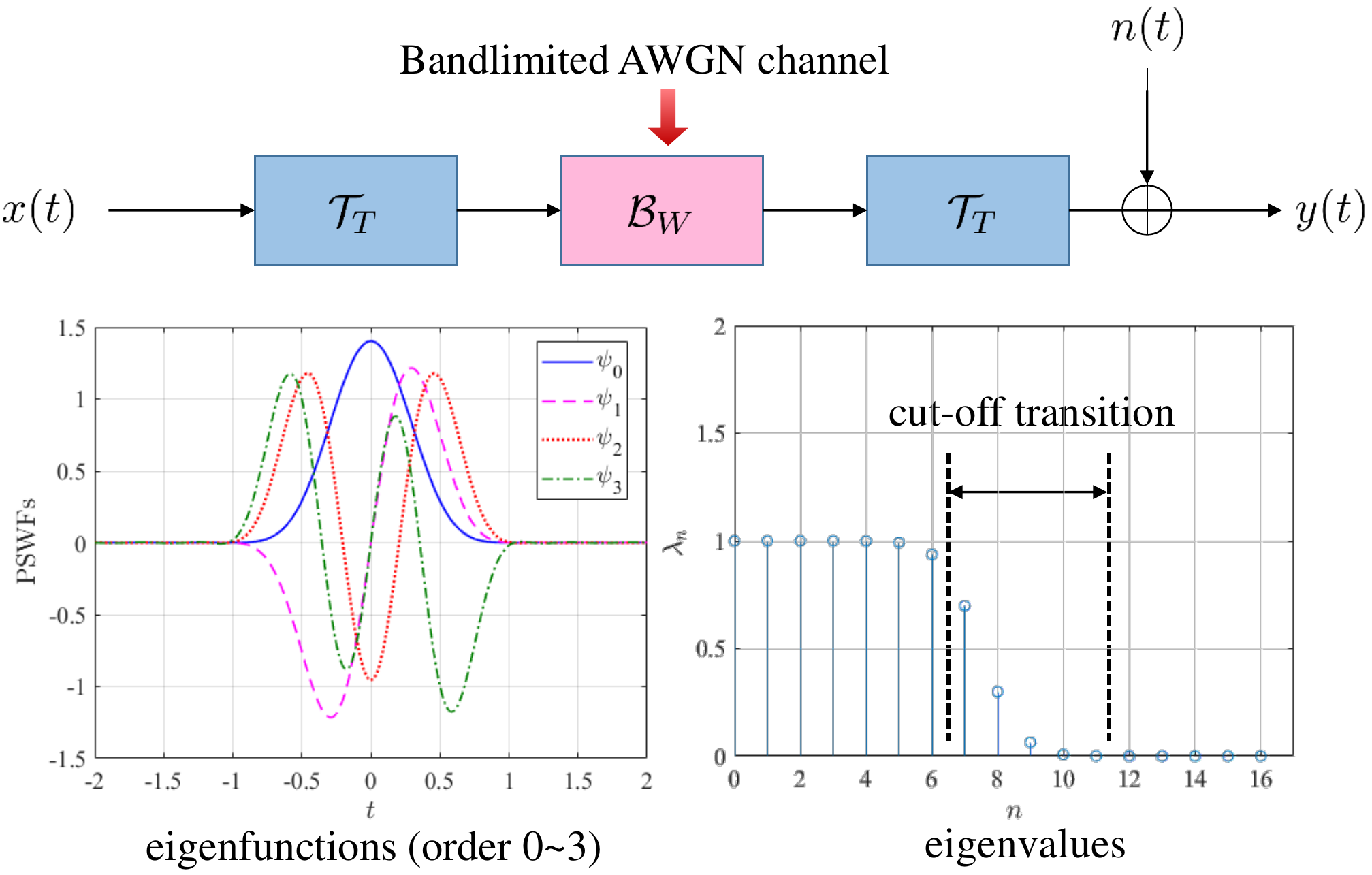} 
	\caption{Example of the functional DoF: The bandlimited AWGN channel and its eigenmodes. The transmitted signal $x(t)$ is time-limited by the temporal truncation operator $\mathcal{T}_T$ before undergoing the $W$-bandlimited channel and a second truncation $\mathcal{T}_T$. The output of this channel $y(t)$ is then obtained by imposing an AWGN process $n(t)$. The eigenmodes of this linear system are given by the prolate spheroidal wave functions (PSWFs) $\{\psi_n\}_{n=0}^{\infty}$, with eigenvalues $\{\lambda_n\}_{n=0}^{\infty}$. In this figure, $T=2\,{\rm s}$, $W=2\,{\rm Hz}$, and thus the functional DoF of the recieved signal $y(t)$ is $2WT=8$. }
	\label{fig:PSWF}
\end{figure}

By contrast, in the information theory community, the DoF has a mathematically rigorous definition that is firmly rooted in the spectral theory of functional analysis, leading to the concept of {\emph{functional DoF}}. 
Generally, if a normed functional space $\mathcal{X}$ contains an $N$-dimensional subspace $\mathcal{X}_N$ such that any function $f\in\mathcal{X}$ can be ``well-approximated'' by some $\hat{f}\in\mathcal{X}_N$, then it is reasonable to claim that the space $\mathcal{X}$ has an essential DoF of $N$.  
An inspiring and important example would be evaluating the DoF of a waveform channel bandlimited to $[-W, W]\, {\rm Hz}$, as is shown in Fig.~\ref{fig:PSWF}. Equivalently, we want to know how many real numbers can be communicated per unit time through this $W$-bandlimited channel. 
Slepian~\cite{slepian1961prolate} solved this problem by collecting all the $W$-bandlimited signals together into a functional space $\mathcal{B}_W$, and asking at least how many coefficients $\{x_n\}_{n=1}^{N_\epsilon}$ are needed to approximate an arbitrary $f(t)\in \mathcal{B}_W$ up to a precision of $\epsilon$, i.e., $\|f(t)-\sum_{n=1}^{N_\epsilon}x_n\phi_n(t)\|/\|f\|\leq \epsilon$.
This minimum number of coefficients $N_\epsilon$ is called the Kolmogorov $N$-width $N_\epsilon(\mathcal{B}_W)$ of the functional space $\mathcal{B}_W$ under given norm $\|\cdot\|$. Since data transmission usually takes place within a finite time interval $[-T/2, T/2]$, the norm is chosen to be $L^2(-T/2, T/2)$. Then, the conclusion is that $N_\epsilon=2WT+\mathcal{O}(\log (WT))$~\cite{slepian1961prolate} holds for any given $\epsilon>0$. 
This means that the $W$-bandlimited channel can essentially transmit $2WT$ real numbers within time $T$. 
In this case, the optimal basis waveforms $\psi_n(t)$ are prolate spheroidal wave functions (PSWFs), which are shown in Fig.~\ref{fig:PSWF}. Note that the functional DoF of the received signal $y(t)$ is determined by evaluating the largest $n$ before the eigenvalues $\lambda_n$ exhibit a cut-off transition behavior.

\vspace{-1em}
\subsection{Channel Capacity}
\label{Sec_2_Subsec_2}
Different from the notion of channel DoF, which characterizes the number of orthogonal subchannels available, the channel capacity measures the error-free information transmission capability of the channel. 
Since the transmission errors are caused by random channel noise, the channel is information-theoretically defined as a conditional transition probability that randomly maps the input to the output of the channel. 
Then, the mutual information between the input and output can be defined, and the channel capacity is derived by taking the supremum of the mutual information over all possible input distributions of the channel. 

The operational meaning of the channel capacity is established by Shannon in his seminal paper~\cite{shannon1948mathematical} in 1948. To establish the operational meaning, he proved the {\it achievability} and {\it converse} theorem of the channel capacity. The {\it achievability} states that, for any given error probability and data rate lower than the capacity, there exists a pair of encoder-decoder of sufficient code length to operate below such an error probability. 
The {\it converse} theorem states that, for an arbitrary data rate higher than capacity, no matter what kind of code is employed, the error probability is bounded away from zero. This prominent result indicates that, it is impossible to realize error-free transmission at a rate higher than the capacity. Thus, the value of channel capacity is established to be the fundamental transmission limit of a given channel. 

\begin{figure}
	\centering 
	\includegraphics[width=0.9\linewidth]{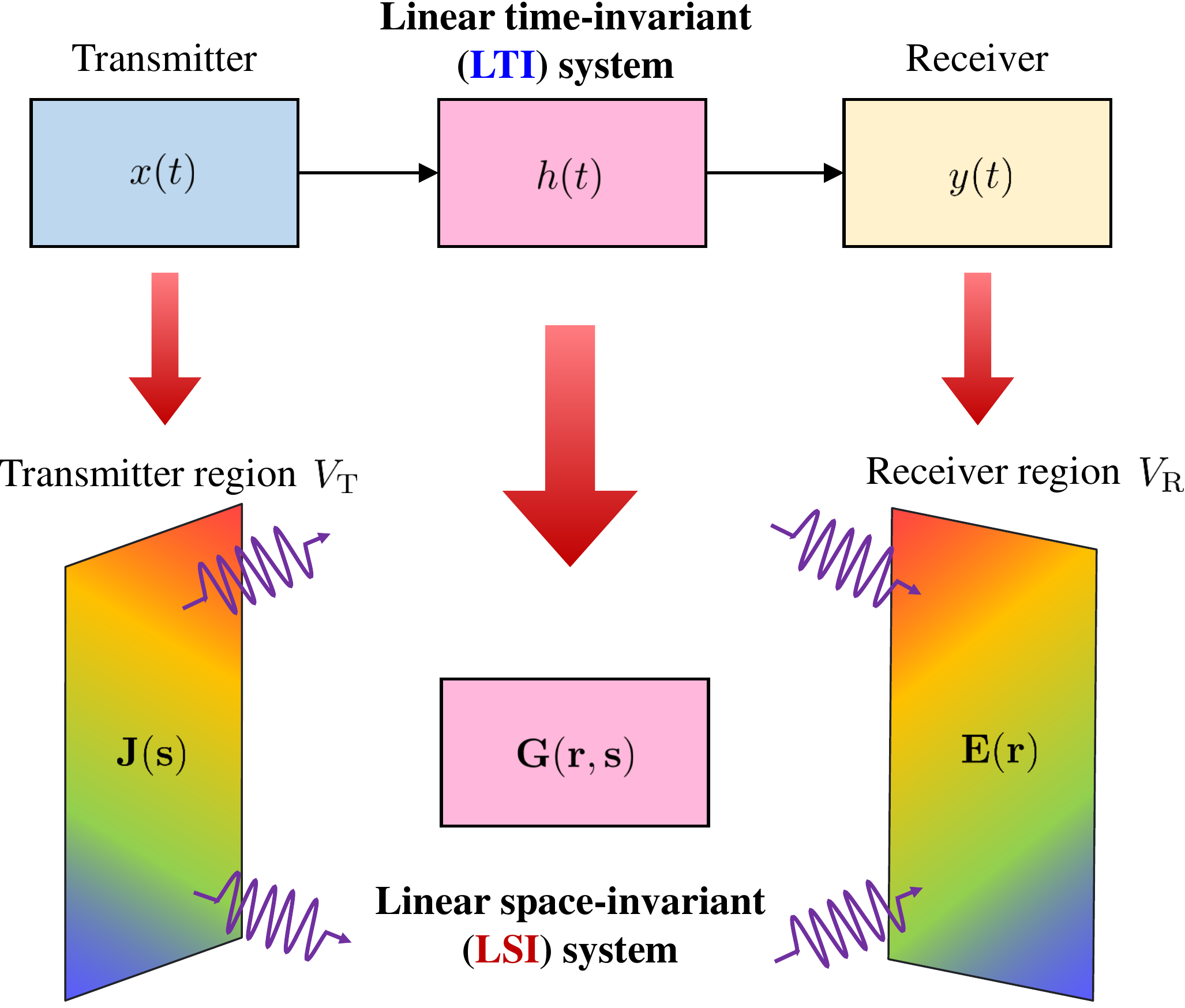} 
	\caption{In analogy with linear time-invariant systems described by a time-domain impulse response $h(t)$ in classical information theory, the EIT model is based on linear space-invariant systems described by the Green's function ${\bm G}({\bm r}, {\bm s})\in\mathbb{C}^{3\times 3}$ from the transmitter region $V_{\rm T}$ to the receiver region $V_{\rm R}$, where spatial coordinates ${\bm s}\in V_{\rm T}$ and ${\bm r}\in V_{\rm R}$ represent the source and the receiver coordinates, respectively.} 
	\label{fig:LTI_LSI}
\end{figure}

\vspace{-1em}
\subsection{Electromagnetic Theory}
\label{Sec_2_Subsec_3}
EM theory is a branch of physics that studies the EM interaction in four-dimensional spacetime from a field-theoretic point of view. The EM forces are carried by EM vector fields, which are usually described by Maxwell's equations. Maxwell's equations are four linear partial differential equations that characterize how the electric fields and magnetic fields are altered by each other and by charges and currents. Combining these four equations, a vector wave equation can be obtained that manifests the existence of EM waves. 

In wireless communications, to describe the electric field ${\bm E}({\bm r})\in\mathbb{C}^3$ at the receiver induced by the source current ${\bm J}({\bm s})\in\mathbb{C}^3$ at the transmitter, the 3D Green's function ${\bm G}({\bm r}, {\bm s})\in\mathbb{C}^{3\times 3}$ is usually introduced as the spatial impulse response of EM systems~\cite{stratton2007electromagnetic}, as is shown in Fig.~\ref{fig:LTI_LSI}. 

Note that EM theory is a deterministic theory built on partial differential equations. Therefore, the full EM response is uniquely determined by the boundary conditions of the EM problem.
As a result, pure EM theory cannot capture the random nature that arises in wireless communications. 

\subsection{Random Fields for EIT}
\label{Sec_2_Subsec_4}
In order to apply information-theoretic analysis to EM fields, probabilistic measures should be assigned to the functional space containing all possible EM fields that satisfy Maxwell's equations, which leads to the random field modeling for EIT. 
Gaussian random processes, usually indexed by the continuous time $t$, are widely applied to model communication signals, from which entropy rates and mutual information formulas can be derived. Gaussian random fields (GRFs) are generalized random processes that allow multiple index variables. Mathematically, GRFs are random multivariate functions whose finite-dimensional marginal distributions are Gaussian. Justified by the central limit theorem, many complicated spatially dependent random values that arise in wireless communications can be modeled by GRFs.  
Inheriting favorable mathematical properties from Gaussian distributions, GRF models exhibit good analytical properties that can facilitate both theoretical deduction and numerical inference. 

In EIT, GRFs can be utilized to model both the EM channels~\cite{marzetta2022fourier} and the EM signals~\cite{wan2022mutual}. 
In real-world communication systems, the EM channel response may vary randomly as a function of spacetime, so it is reasonable to model such spatio-temporal variations by a channel GRF. 
This allows the Bayesian inference of channel entries that are not directly measured. 
Similarly, in order to capture the randomness of the transmitted and received signals, the current distributions at the transmitter and the electric fields at the receiver can also be modeled by GRFs~\cite{wan2022mutual}, from which the information-theoretic mutual information can be defined to characterize the information transmission capability of such EM channels. The GRF-based information-theoretic analysis will be explained with detail in Section~\ref{sec_3_subsec_3}.

\section{Modeling Methodologies for EIT}
In this section, we discuss the continuous modeling schemes for EIT, which include both continuous channel modeling and noise field modeling. 

\vspace{-1em}
\subsection{Continuous Channel Modeling}
In this part, we will discuss the channel models in EIT. The basic idea is to describe the EM channel by a bounded linear operator $T$ that maps the source current distribution ${\bm J}({\bm s})\in\mathcal{L}^2(V_{\rm T})$ to the noiseless received field ${\bm E}({\bm r})\in\mathcal{L}^2(V_{\rm R})$.  
Note that to model a traditional narrowband MIMO channel, a matrix with complex-valued entries ${\bm H}:\mathbb{C}^M\to\mathbb{C}^N$ is usually utilized, where each entry represents the complex channel between a pair of transceiver antennas. 
This channel model is, in essence, spatially discrete. 
However, in EIT, it is reasonably assumed that one can measure the field at an arbitrary point inside the receiver region, where the precision of such a measurement is subject to some physical noise limits. This leads to the assumption of EIT that the transceivers operate in a continuous functional space $\mathcal{L}^2(V_{\{{\rm T, R}\}})$.  

To ensure compatibility with discrete MIMO channel modeling, the connection between the entries of ${\bm H}$ and the transceiver antenna modes are usually assumed to be ${\bm H}_{qp}=\langle\varphi_q|T|\phi_p \rangle$, where $\phi_p$ and $\varphi_q$ are the $p$-th and $q$-th operating modes of the transmit and receive antennas, respectively. Thus, modeling the EM channel by a bounded linear operator $T$ is mathematically consistent with the existing matrix modeling. 

The EIT channel can either be deterministically modeled or stochastically modeled. Deterministic channel models have simpler mathematical expressions, but stochastic models usually better capture the small-scale fast fading caused by multipath effects and user mobility. 
The simplest deterministic channel model is the line-of-sight (LoS) model, which assumes that there is only one direct link without any scatterers between the transceivers. In this scenario, the channel operator is the free-space Green's function ${\bm G}({\bm r}, {\bm s})$. 

\begin{figure}
	\centering 
	\includegraphics[width=\linewidth]{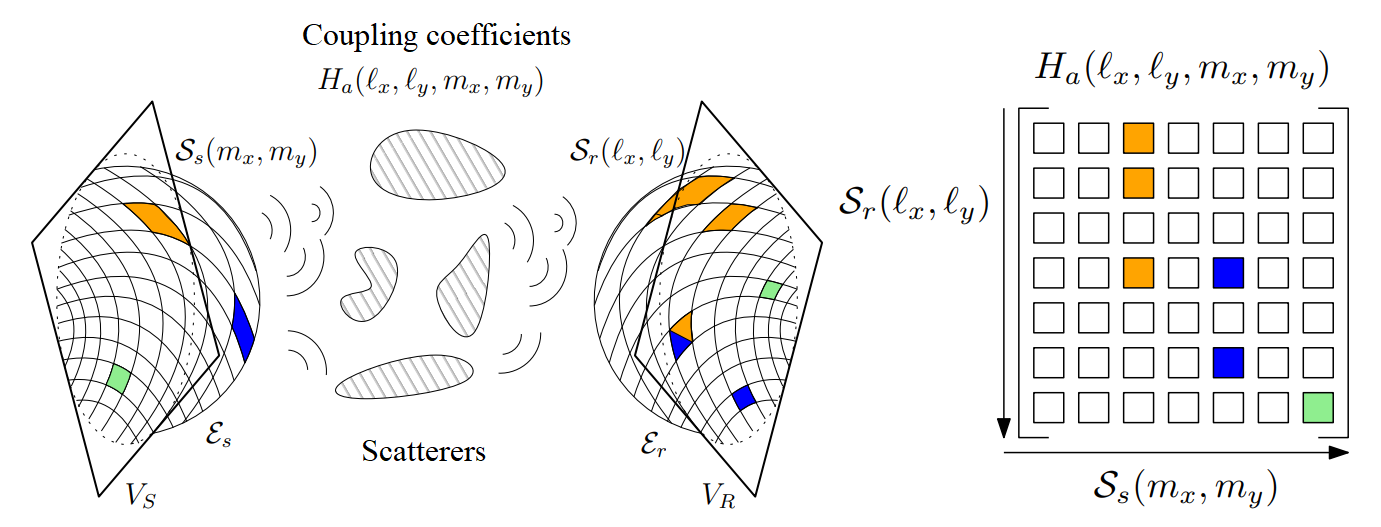} 
	\caption{Continuous stochastic channel modeling, where the linear channel operator is projected onto the Fourier plane-wave basis~\cite{marzetta2022fourier}.} 
	\label{fig:marzetta}
\end{figure}
Another approach is stochastic channel modeling, which usually specifies the autocorrelation function ${\bm R}({\bm r}, {\bm s}; {\bm r}', {\bm s'})$ between the ${\bm s}\to {\bm r}$ channel and the ${\bm s}'\to{\bm r}'$ channel coefficients. As introduced in Subsection~\ref{Sec_2_Subsec_4}, this autocorrelation function determines a GRF that enables useful tools such as Bayesian inference for channel estimation. 
For stochastic channel modeling, a recent novel method is named as Fourier plane-wave expansion (see Fig.~\ref{fig:marzetta}), where the channel is modeled by a spatial GRF which is provably compatible with the EM Helmholtz equation~\cite{marzetta2022fourier}. This EM-compatible modeling enables further information-theoretic analysis that reveals the fundamental limit of electromagnetic laws on information transmission. 

\vspace{-3mm}
\subsection{Noise Field Modeling}
After the discussion of the continuous random channel models, we will introduce the noise modeling scheme. 
Noise is vital in information theory, because it is a key factor to determining the capacity. 
In classical information theory, the noise is usually modeled by a time-domain additive white Gaussian noise (AWGN) with a constant power spectral density $n_0/2$. Thus, its projection coefficients onto any orthonormal basis are independent and of equal power $n_0/2$, which simplifies theoretical analysis. 
Similarly, in EIT, the noise is usually modeled as a spatial AWGN with flat wavenumber PSD. This spatial AWGN implies that the additive noises at any disjoint small spatial regions $V_1, V_2$ are independent and identically distributed (i.i.d.) complex Gaussian random variables, whose variances are proportional to the volumes $\mu(V_1), \mu(V_2)$ of the small regions. Although the AWGN model facilitates theoretical analysis, the white spectral assumption is problematic, since it causes an unbounded noise power. 


\section{Information-Theoretic Analysis for EIT}
In this section, based on the EIT modeling methodologies above, we discuss the basic performance indicators and the corresponding analysis techniques of EIT, including the functional DoF, the channel DoF, and the mutual information.  

\vspace{-1em}
\subsection{Functional DoF}\label{Sec_4_Subsec_1}
As we have clarified in Section~\ref{Sec_2_Subsec_2}, the functional DoF in classical information theory represents the minimum number of required samples to reconstruct the signal. Similarly, in EIT, the functional DoF refers to the minimum number of required samples $N_\epsilon$ to reconstruct a given EM field. 
Such a functional DoF is closely related to the transmission capability of the EM system, because the minimum number of required samples to reconstruct an EM field is equivalent to the maximum number of complex values that can be transmitted within a single EM channel use. 

The functional DoF of a scattered EM field ${\bm E}({\bm r})$ can be analyzed in the transform domain ${\bm E}({\bm k}) = \mathcal{F}^3[{\bm E}({\bm r})]({\bm k})$, i.e., the wavenumber domain. Resembling the bandlimited signals in the time-frequency domain, the radiated EM fields also exhibit the wavenumber-limited property in the space-wavenumber domain. 
Thus, it can be easily proved that a half-wavelength sampling suffices to asymptotically reconstruct an arbitrary EM field up to any given precision $\epsilon$ as the receiving region $\mu(V_{\rm R})\to\infty$, i.e., the EM DoF is at most proportional to the number of half-wavelength grids in the receiver region. 

A more refined analysis of the wavenumber-limitedness of the EM fields shows that the EM functional DoF depends on how rapidly the phase of a radiated field changes over space. 
To describe such a phase change, the authors of~\cite{bucci1987spatial} introduced the spatial bandwidth $W$ to describe the degree of wavenumber-limitedness. 
It is proved that for electromagnetic sources confined to a sphere of radius $a$, the spatial bandwidth satisfies $W\leq \sqrt{2}\beta a$, where $\beta$ is the propagation constant of the time-harmonic EM fields. 
The functional DoF of the received field is thus at most proportional to $W$ and the length of the observation region. 


\vspace{-1em}
\subsection{Channel DoF}
The functional DoF introduced in the previous subsection emphasizes the intrinsic DoF of the received field. 
However, due to the restrictions of the EM channel, it is possible that some of these functional DoFs at the receiver cannot be excited, especially in low-rank LoS propagation conditions. Thus, evaluating the channel DoF is of practical importance.  

As is defined in Subsection~\ref{Sec_2_Subsec_1}, the channel DoF represents the maximum number of independent parallel channels that can be used to transmit information. 
Similar to the SVD decomposition of matrices, the DoF of a LoS EM channel can be solved by expanding the channel operator $T$ onto a series of orthogonal sub-channels: $T=\sum_n \sigma_n |\varphi_n\rangle\langle\phi_n|$, and counting the number of channel gains $\sigma_n$ that exceed a certain threshold. 
Specifically, in the scenario where a pair of coaxial square transceivers are employed as transmit and receive antennas, the LoS channel operator $T$ is given by the free-space Green's function ${\bm G}({\bm r}, {\bm s})$, and the corresponding eigenvalue problem can be approximated by the standard Slepian's concentration problem~\cite{miller2000communicating}. In this case, the channel eigenmodes $\varphi_n$ and $\phi_n$ are proved to be well-approximated by prolate spheroidal wave functions $\psi_n$.   
Through this approach, the DoF of LoS channel model has been approximately derived to be proportional to the product of the area of the transceivers~\cite{pizzo2022nyquist,miller2000communicating}. 

For non-line-of-sight (NLoS) deterministic channel models, the channel DoF can be similarly derived by performing Slepian's analysis in the angular domain, since the scatterers usually appear in limited solid angular regions. 
Following Slepian's analysis, it is proved that the NLoS channel DoF is proportional to both the solid angle of the scatterer cluster and the geometric lengths of the transceivers~\cite{poon2005degrees}.   
If the channel is modeled as a stochastic channel with random scatterer positions, von Mises-Fisher distributions~\cite{byers2004spatially} can be used to model the statistical characteristics of the channel. 
The channel DoF in this stochastic model should be the expected value averaged by probability.

\begin{figure}
	\centering
	\includegraphics[width=1.0\linewidth]{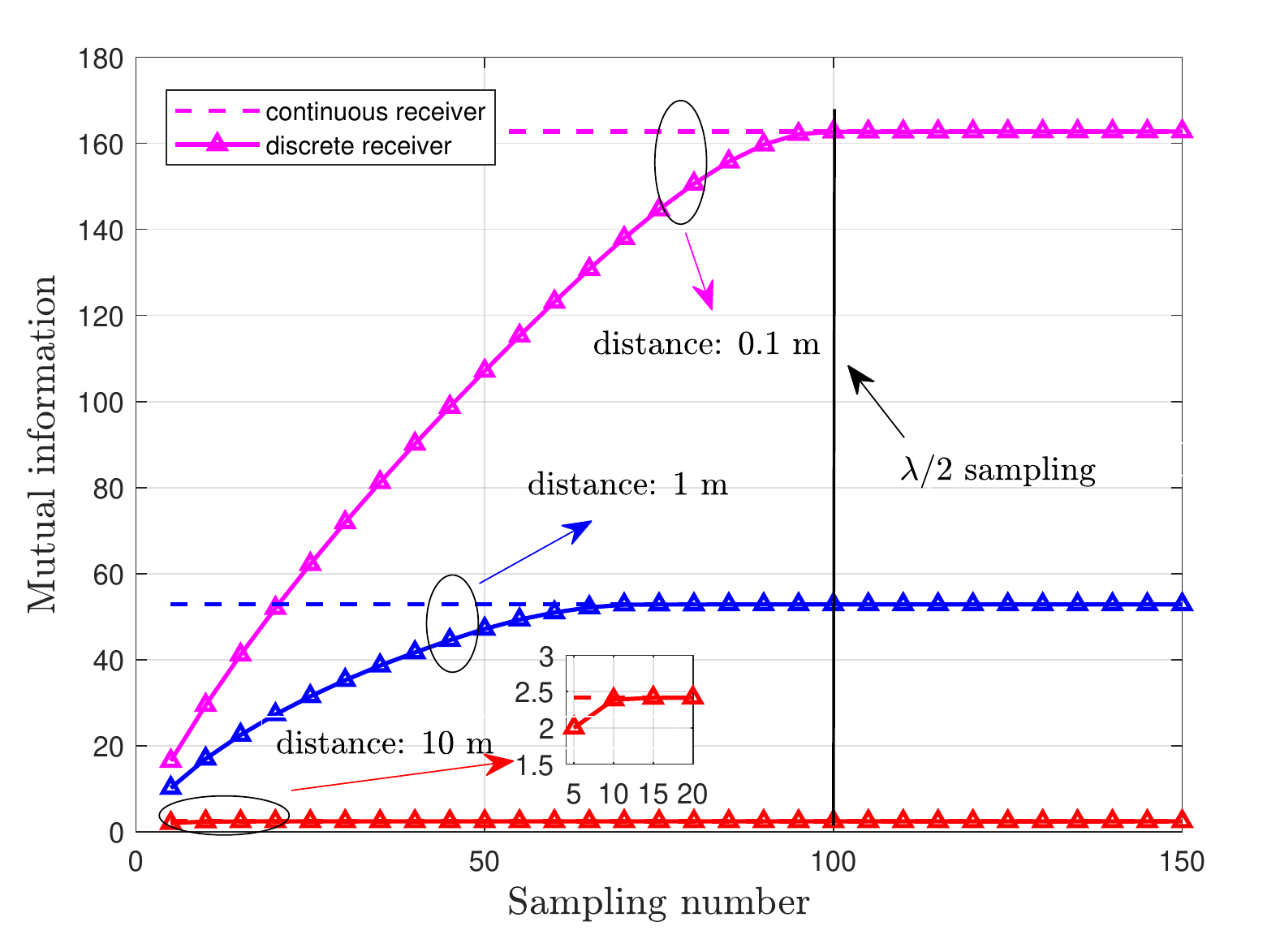}
	\caption{Convergence of the discrete MIMO mutual information to the continuous-space EIT mutual information $I({\bm J}; {\bm E})$. The $x$-axis (sampling number) represents the number of antennas placed in the receiver region $V_{\rm R}$. }
	\label{fig:MIMO_EIT_comparison}
\end{figure}

\vspace{-3mm}
\subsection{Mutual Information Analysis} \label{sec_3_subsec_3}
Besides the DoF which reveals the number of available sub-channels, the channel mutual information $I({\bm J}; {\bm E})$ is another important performance indicator. 
Compared to the traditional MIMO information theory using random vectors as transmitted and received signals, EIT treats the EM fields ${\bm J}, {\bm E}$ as GRFs. 
The random field modeling follows the statistical approach of mutual information analysis by Shannon, and can be viewed as a continuous extension of the traditional analysis based on random vectors. 
In parallel with MIMO theory based on matrices, operator theory can be used to describe the autocorrelation ${\bm R}({\bm r}, {\bm r}')=\mathbb{E}[{{\bm E}({\bm r}){\bm E}\H ({\bm r}')}]$ of the random field. 
Different from the matrix determinant form of MIMO mutual information, the EIT information takes a Fredholm determinant form $\log\det({\bf I}+T_{\bm E}T_{\bm N}^{-1})$~\cite{wan2022mutual}, where $T_{\bm E}$ and $T_{\bm N}$ are self-adjoint autocorrelation operators of the random fields ${\bm E}({\bm r})$, and ${\bm N}({\bm r})$, respectively.  
This Fredholm determinant form provides a closed-form formula for the EIT mutual information. Thus, numerical schemes for evaluating Fredholm determinants can be transplanted onto the computation of this EIT mutual information, as is shown in Fig.~\ref{fig:MIMO_EIT_comparison}. It is numerically verified that the discrete MIMO mutual information converges to the continuous-space EIT mutual information, which justifies the Fredholm definition of the EIT mutual information. 

\section{Applications of EIT}
In this section, we present several EIT-inspired applications. 
These applications either mimic the analysis methodologies of EIT to fully achieve the existing DoF, or try to explore new communication resources predicted by EIT. 

\vspace{-1em}
\subsection{Continuous Aperture MIMO (CAP-MIMO)}
Traditional wireless communication systems deploy finite antennas in a limited aperture at the transceivers. 
However, the EIT mutual information analysis is built on spatially continuous theoretical frameworks. 
To bridge the gap between the performance limit in MIMO theory and that in EIT, CAP-MIMO, also known as holographic MIMO, has attracted increasing research interests recently~\cite{zhang2022pdma}. 
CAP-MIMO adopts a hypothetical structure that contains infinitely dense antennas in a limited spatial region, which is capable of continuously generating arbitrary current distributions at the transmitter and detecting arbitrary electric fields at the receiver.  
The current distributions on the transmitter are called patterns of the CAP-MIMO. 
Different signals are modulated onto different patterns before radiating into the space. 
These continuous patterns need to be optimized by specially-designed algorithms in order to achieve a higher multi-user data rate~\cite{zhang2022pdma}.


\subsection{Location Division Multiple Access (LDMA)}
\begin{figure*}[t]
	\centering 
	\includegraphics[width=0.9\textwidth]{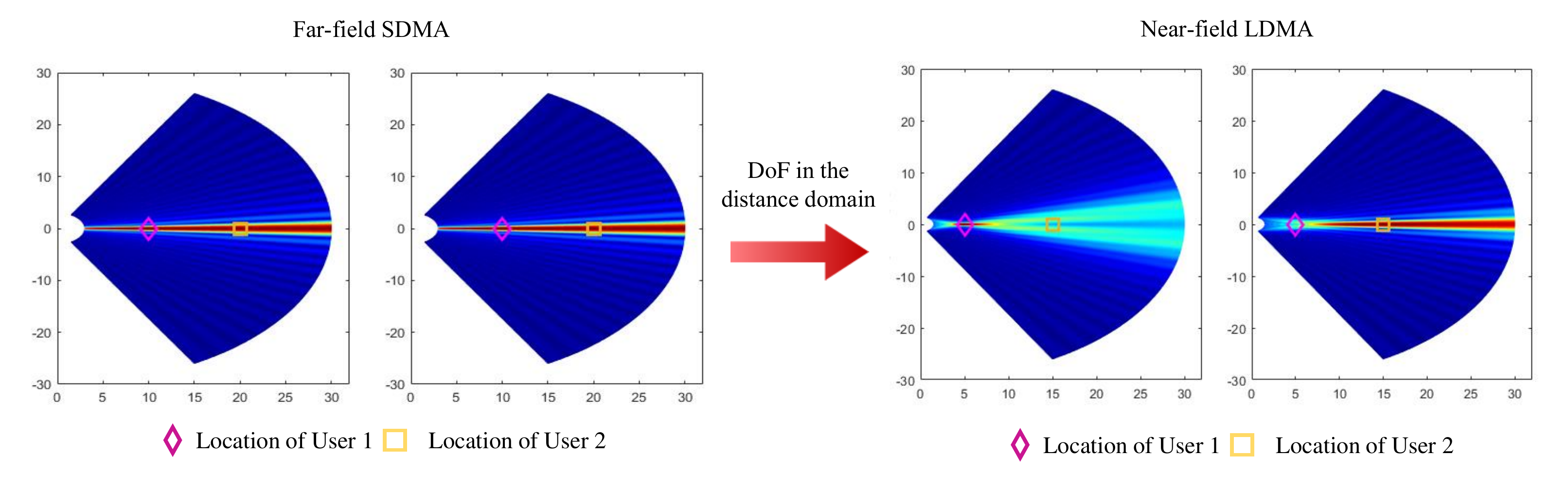} 
	\caption{Comparison between far-field SDMA and near-field LDMA, where the DoF in the distance domain can be exploited to provide a new possibility for capacity improvement~\cite{wu2022multiple}. }
	\label{fig:LDMA}
\end{figure*}

Traditional far-field spatial division multiple access (SDMA) scheme exploits the angular orthogonality of the far-field planar-wave propagation channel to serve users from different angles simultaneously. In order to further improve the communication rate, extremely large antenna arrays (ELAAs) are introduced for improving the spectral efficiency. Different from the traditional far-field propagation environment, this enlarged array aperture will inevitably introduce the {\it near-field} spherical wave propagation characteristics.  

Although the near-field effect seems to corrupt the planar wave assumptions and make the traditional DFT codebook-based angular domain beamforming techniques no longer applicable, it also brings new potential for data transmission and multiplexing. This is because the more complicated near-field propagation channel exhibits additional orthogonality apart from the already well-known angular orthogonality in SDMA-based systems. 
The additional orthogonality comes from the distance domain, i.e., the channels of users located at different distances are proved to be orthogonal in the near-field region~\cite{wu2022multiple}, thus providing extra DoFs in the distance domain. Thus, by re-designing the near-field codebook to match the near-field propagation characteristics of the EM waves, a new multiple access scheme called the location division multiple access (LDMA) is proposed to enhance the overall multiple access capability of the communication systems.

\section{Open Problems for EIT}
In this section, we will discuss some open problems for EIT. We will also explain the meanings of these open problems, and why they are important for the future research of EIT. 

\subsection{Digital and Analog Challenges}
To achieve the capacity improvement predicted by EIT, transceivers are expected to employ fully-continuous antenna surfaces. However, the implementation of such spatially continuous antenna surfaces will inevitably require the processing of an enormous volume of baseband data and the deployment of a vast amount of electrically small antenna patches. Thus, the exploding number of antennas will bring fundamentally new challenges to both the digital and analog signal processing. 

In the digital domain, the optimal current distribution patterns should be frequently updated within a typical timescale of 1\,ms to accommodate the fast-varying EIT channel. In addition, the DFT bases have to be further improved since the EIT eigenmodes are generally not sinusoidal~\cite{marzetta2022fourier}. Thus, numerous nearly-continuous bases have to be frequently designed in the digital domain, which is a great challenge even to the state-of-the-art baseband processors.  

In the analog domain, there exists a physical tradeoff between the performance and the size of antennas~\cite{stratton2007electromagnetic}. Thus, it is practically difficult to design near-continuous antennas without sacrificing the radiation efficiency and working bandwidth. Apart from the physical tradeoff, the mutual coupling between adjacent antennas need to be carefully suppressed when the antenna array becomes denser, in order not to churn the designed radiation patterns and degrade the communication performance.

\subsection{Compatible Noise Modeling in Discrete and Continuous Communication Systems}
It can be proved that, the purely i.i.d. thermal noise distribution will cause the divergence of the end-to-end capacity of a MIMO transceiver, when the number of receiving antennas increases indefinitely within a constrained aperture size. 
The reason is that, when the number of receiving antennas increases, the signals can be aggregated coherently within any small spatial region, while the noises add up non-coherently because they are uncorrelated. 
Thus, in this small region, the signal energy scales quadratically, while the noise energy scales linearly, resulting in an unbounded linear improvement of the signal-to-noise ratio (SNR). Thus, the capacity will diverge to infinity, contradicting the principle of energy conservation.  

This absurdity is, in fact, caused by the improper assumption that the noise is spatially uncorrelated. 
If we assume a correlated noise, then this capacity divergence naturally disappears. 
As a result, correlated noise models are required for the EIT capacity analysis. On one hand, the noise model may be tuned to exhibit a small enough spatial ``coherence length'' to be compatible with the independent MIMO noise model with the half wavelength-spaced antennas. On the other hand, the noise should possess some spatial correlation on a small scale to ensure a finite-valued EIT capacity. 
The construction of such a kind of noise model that bridges the macro-scale and micro-scale noises is, up to date, an open problem. 



\subsection{Evaluation of the EM Capacity}
In classical information theory, the ``capacity'' is defined as the supremum of all the operationally achievable transmission rates. It is favorable that in discrete memoryless channels, the capacity equals the maximum mutual information. However, this conclusion has not been proved for continuous EM waveform channels. In the existing works on EIT~\cite{wan2022mutual,zhang2022pdma,marzetta2022fourier}, maximum EIT mutual information values are calculated under the assumptions of linear deterministic EM channels and continuous transceivers, and the EIT mutual information is further compared to the discrete MIMO mutual information. Unfortunately, these EIT mutual information values only serve as an upper bound to the EM operational capacity. 
Thus, in the strict sense, the EM capacity is still an open problem. 

\section{Conclusions}
In this paper, we have investigated the fundamental mathematical tools, basic modeling methodologies, and theoretical performance analyses that constitute EIT. 
Furthermore, we have discussed some recent novel applications related to EIT, aiming at designing new wireless communication systems for single-user and multi-user capacity enhancement. 
The recent progress on EIT has demonstrated its potential to become a unified and widely applicable theory for the EM physical layer. However, there are still some unresolved open problems that require further study in the future. 

\footnotesize

\bibliographystyle{IEEEtran}
\bibliography{EIT, IEEEabrv}

\normalsize
\vspace{-1em}
\section*{Biographies}

{\bf Jieao Zhu} is a Ph.D. student from the Department of Electronic Engineering at Tsinghua University, Beijing, China. His research interests include electromagnetic information theory (EIT), coding theory, and quantum computing. 
\\

{\bf Zhongzhichao Wan} is a Ph.D. student from the Department of Electronic Engineering at Tsinghua University, Beijing, China. His research interests include electromagnetic information theory (EIT), coding theory, and channel modeling. 
\\

{\bf Linglong Dai} is a Professor from Tsinghua University. His current research interests include massive MIMO, RIS, Wireless AI, and EIT. He has received the IEEE ComSoc Leonard G. Abraham Prize in 2020, the IEEE ComSoc Stephen O. Rice Prize in 2022, and the IEEE ICC 2022 Outstanding Demo Award. He was listed as a Highly Cited Researcher by Clarivate from 2020 to 2022. He was elevated as an IEEE Fellow in 2022.
\\

{\bf M\'{e}rouane Debbah} is Professor at Khalifa University of Science and Technology in Abu Dhabi. From 2014 to 2021, he was vice-president of the Huawei France Research Center, where he was jointly the director of the Mathematical and Algorithmic Sciences Lab as well as the director of the Lagrange Mathematical and Computing Research Center.
\\

{\bf H. Vincent Poor} is the Michael Henry Strater University Professor at Princeton University, where his interests include information theory, machine learning and network science, and their applications in wireless networks, energy systems and related fields. Among his publications in these areas is the recent book Machine Learning and Wireless Communications (Cambridge University Press, 2022). A member of the U.S. National Academies of Engineering and Sciences, he received the IEEE Alexander Graham Bell Medal in 2017. 
\\

\end{document}